\documentclass[%
 aps,
 amsmath,amssymb,
 reprint,%
]{revtex4-1}

\usepackage{graphicx}% Include figure files
\usepackage{dcolumn}% Align table columns on decimal point
\usepackage{bm}% bold math
\usepackage{url}
\usepackage{ulem}
\usepackage{color}
\definecolor{dred}{rgb}{.8,0.2,.2}

\usepackage[utf8]{inputenc}
\usepackage[T1]{fontenc}
\usepackage{mathptmx}
\usepackage{etoolbox}

\makeatletter
\def\@email#1#2{%
 \endgroup
 \patchcmd{\titleblock@produce}
  {\frontmatter@RRAPformat}
  {\frontmatter@RRAPformat{\produce@RRAP{*#1\href{mailto:#2}{#2}}}\frontmatter@RRAPformat}
  {}{}
}%
\makeatother
\begin{document}

\preprint{AIP/123-QED}

\title{Mixed-signal data acquisition system for optically detected magnetic resonance of solid-state spins}
% Force line breaks with \\
\author{Feifei Zhou}
\affiliation{School of Physics, Hefei University of Technology, Hefei, Anhui 230009, China}
 
\author{Shupei Song}
\affiliation{School of Physics, Hefei University of Technology, Hefei, Anhui 230009, China}

\author{Yuxuan Deng}
\affiliation{School of Physics, Hefei University of Technology, Hefei, Anhui 230009, China}

\author{Ting Zhang}
 \altaffiliation{zhangting@hfut.edu.cn}
\affiliation{School of Physics, Hefei University of Technology, Hefei, Anhui 230009, China}

\author{Bing Chen}
\affiliation{School of Physics, Hefei University of Technology, Hefei, Anhui 230009, China}

\author{Nanyang Xu}
 \altaffiliation{Author to whom correspondence should be addressed: nyxu@hfut.edu.cn}
\affiliation{School of Physics, Hefei University of Technology, Hefei, Anhui 230009, China}

\date{\today}

\begin{abstract}
We report a mixed-signal data acquisition (DAQ) system for optically detected magnetic resonance (ODMR) of solid-state spins. This system is designed and implemented based on a Field-Programmable-Gate-Array (FPGA) chip assisted with high-speed peripherals. The ODMR experiments often require high-speed mixed-signal data acquisition and processing for general and specific tasks. To this end, we realized a mixed-signal DAQ system which can acquire both analog and digital signals with precise hardware synchronization. The system consist of 4 analog channels (2 inputs and 2 outputs) and 16 optional digital channels works at up to 125 MHz clock rate. With this system, we performed general-purpose ODMR and advanced Lock-in detection experiments of nitrogen-vacancy (NV) centers, and the reported DAQ system shows excellent performance in both single and ensemble spin cases. This work provides a uniform DAQ solution for NV center quantum control system and could be easily extended to other spin-based systems.
\end{abstract}

\maketitle

\section{Introduction}
With the development of quantum information science, various quantum systems have widely potential applications in the rapidly emerging research fields like quantum computation\cite{suterCOM, neumannCOM, kongfeiCOM, ionsCOM, atomCOM}, quantum simulation\cite{wuSIM, ionsSIM, ywSIM, atomSIM, chenSIM}, and quantum metrology\cite{Msensing1, atomsensing, ionssensing, SQUIDsensing}. In particular, the nitrogen-vacancy (NV) centers in diamond work as excellent quantum sensors for magnetic fields\cite{Msensing2, Msensing3}, electric fields\cite{Esensing}, and temperatures\cite{Tsensing}. With the assistance of optically detected magnetic resonance (ODMR) technique, manipulation and detection of NV centers can be easily realized by utilizing laser pumping and resonant alternating magnetic fields\cite{NVcenter}, where the spin-state-dependent fluorescence can be collected by suitable detectors depending on its intensity. Emitting fluorescence of single spin and small spin ensembles is generally collected by avalanche photo diodes (APD) and transfered into digital pulses. As for large spin ensembles, photodetectors (PD) are suitable for the detection of its strong fluorescence, which transfer the fluorescence into analog signals. For further handling of the information, data acquisition (DAQ) systems with signal sampling, pre-processing, and transmission functions are required in applications, which are widely used in scientific research areas\cite{daq1, daq2, daq3, daq4, daq5, daq6}.

In the field of quantum experiment based on NV centers, especially the single-spin case, DAQ systems from the National Instruments (NI) company are generally adopted, which provide multi-function modules and uniform communication firmwares with computers. Nevertheless, NI DAQ equipments are lack of abilities of high-speed sampling over 10 million samples per second (MSPS). Some other companies like the Spectrum could provide high-speed sampling solutions which are extremely costly and often used in atomic experiment\cite{atomsensing}. Besides, in the NV center experiment, synchronized acquisition of both analog and digital signals are required, which could not be easily realized by the above commercial equipments\cite{NVzongshu}.

\begin{figure*}
	\includegraphics[width=1.8\columnwidth]{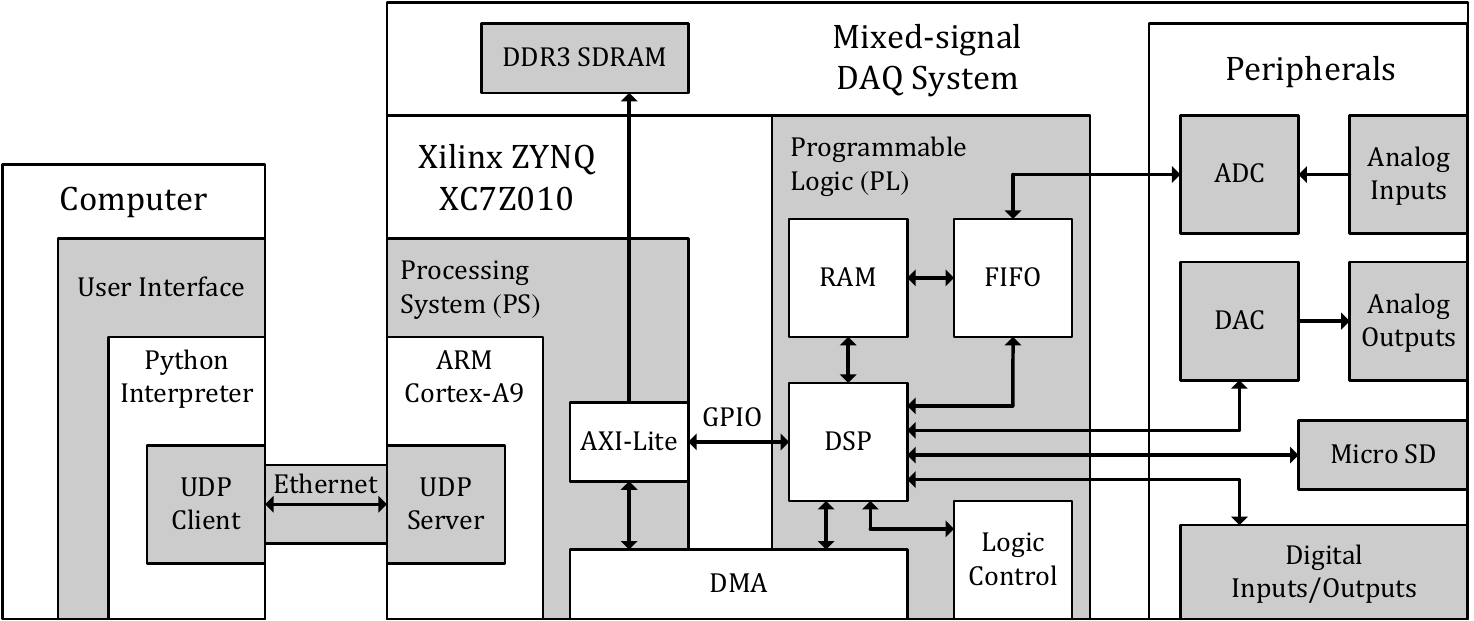}
	\caption{\label{fig1} The architecture of the reported mixed-signal DAQ system based on the FPGA system with several peripherals included. Data transmission between PL and PS is realized by utilizing the DMA, while control commands from PS are transmitted to PL through general-purpose input/output (GPIO) bus. The communication between computer and the mixed-signal DAQ system is established via gigabit ethernet based on UDP protocol, and the customized software is developed in Python.}
\end{figure*}

In this paper, we report a mixed-signal DAQ system based on a Field-Programmable-Gate-Array (FPGA) chip assisted with analog-to-digital converter (ADC) and digital-to-analog converter (DAC) chips. Our DAQ system is capable of handling both analog and digital signals with the sampling rate up to 125 MSPS in synchronization. With this DAQ system, general-purpose experiments of NV centers are realized in both single and ensemble spin cases. We also demonstrate advanced functions of the DAQ system, by which the analog signals ($e.g.$, the PD output) and fluorescence photon counts are synchronously processed. This enables some interesting applications in NV centers, for example, the Lock-in detection and the synchronized ODMR scanning of NV centers. The reported DAQ system can be easily extended to other spin-based systems and could also be a general low-cost solution in a wide-range applications.

\section{Implementation}
The architecture of the reported mixed-signal DAQ system is shown in Fig.~\ref{fig1}. The FPGA-based system mainly consist of a processor (ARM Cortex A9), a FPGA (Xilinx Zynq XC7Z010) chip, 14-bit dual ADC (LTC2145CUP-14) and DAC (AD9767) chips with the sampling rate up to 125 MSPS, 4 analog channels and 16 optional digital channels. There are three main modules customized and embedded in the system: synchronized acquisition and processing (SAP), multiplex signal generator (MSG) and high-speed communication (HSC), which are designed and implemented based on the framework of Zynq XC7Z010 FPGA consisted of programmable logic (PL) and processing system (PS).

\begin{figure}
	\includegraphics[width=1\columnwidth]{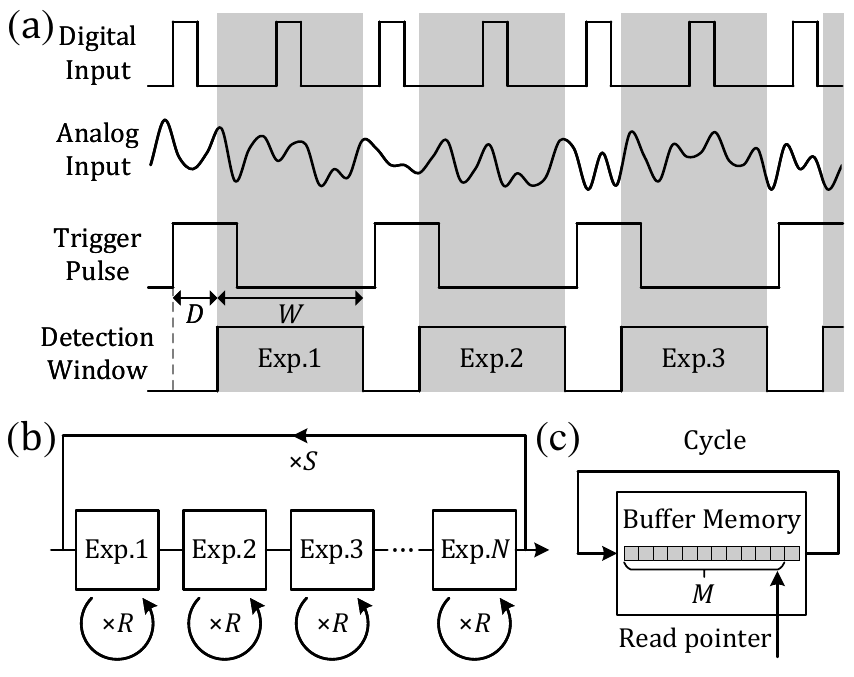}
	\caption{\label{fig2} (a) The timing sequence of synchronized acquisition in SAP module. Both maximum values of two adjustable parameters $D$ and $W$ are $2^{35}$ ns. (b) Schematics of the data processing in SAP module based on sequence pattern. $N$ experimental points can be accomplished point by point in a specific time order determined by parameters $S$ and $R$. (c) Schematics of the data processing in SAP module based on continuous pattern. Here the maximum value of parameter $M$ is 4096.}
\end{figure}

Two analog input (AI) channels and three digital input (DI) channels are configured for the implementation of SAP. Two of DI channels are utilized to be interfaces of external individual trigger events for two AI channels, and the third DI channel is designed for the counting of digital signals in synchronization with one of AI channels. The timing sequence of SAP module is shown in Fig.~\ref{fig2}(a). Two adjustable parameters $D$ and $W$ represent delay time and duration of detection window respectively. When receiving internal or external trigger events, analog signals from two AI channels within $W$ duration can be independently sampled at 8 ns intervals by ADC and cached into the first-in-first-out (FIFO) buffer memory, meahwhile digital signals within $W$ duration from the DI channel are counted by digital-signal-processor (DSP) integrated in PL. Immediately, analog data from FIFO memory and digital data are synchronously processed in a specific pattern and encoded as 128-bit packets by utilizing a mount of DSP slices, then transmitted and storaged into the external double-data-rate three synchronous dynamic random access memory (DDR3 SDRAM) with 512 Megabyte (MB) capacity through the direct memory access (DMA) and advanced extensible interface lite (AXI-Lite) bus. Here logic control of DSP is implemented based on the FPGA logic resources. Finally, high-speed communication between the DAQ system and computer is realized via gigabit ethernet based on user datagram protocol (UDP)\cite{udp}. The client and server of UDP protocol are established by computer and processor in the DAQ system respectively, and the user interface on computer is developed in Python.

\begin{figure*}
	\includegraphics[width=1.6\columnwidth]{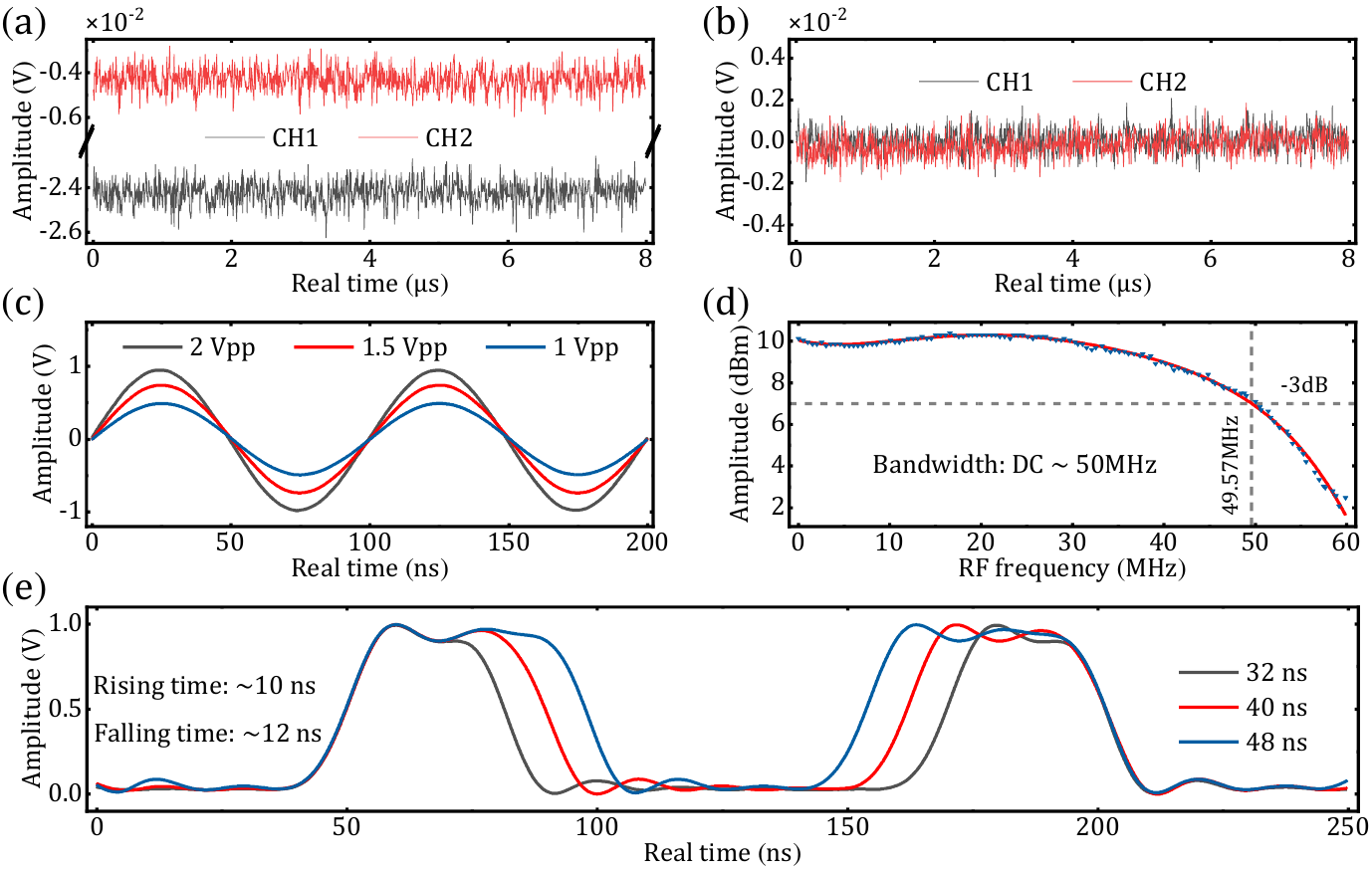}
	\caption{\label{fig3} (a) The bias noise of AI channel 1 (CH1) and channel 2 (CH2) without correction. (b) The bias noise of CH1 and CH2 with correction. (c) The oscillagram of sine-type RF signals output from the DAQ system with the frequency of 10 MHz and adjustable amplitude. (d) Power calibration of RF signals with the frequency range from 0 to 62.5 MHz. (e) The oscillagram of PWM-based digital pulses with adjustable period and duty cycle. (c), (d), and (e) are obtained by a commercial oscilloscope (RIGOL MSO8104).}
\end{figure*}

To satisfy multiple experimental requirements, there are two working patterns customized for the data processing in SAP module: sequence pattern and continuous pattern. To improve the signal-to-noise ratios, general quantum experiments should be repeated for many times in sequence. For this purpose, the sequence pattern is designed and implemented. The sequence pattern allows the acquisition of $N$ experimental points in a specific time order determined by two adjustable parameter $S$ and $R$, the mechanism of which is shown in Fig.~\ref{fig2}(b). $S$ and $R$ represent the repeat times of the whole experimental process and individual experimental point respectively, and the priority of $R$ is higher than one of $S$. Moreover, uninterrupted measurement in real-time domain is equally important in some specific applications. To this end, the continuous pattern is designed by configuring a buffer memory with 64 Kilobyte (KB) capacity in block random access memory (BRAM) of FPGA. When the continuous pattern is enabled, samples are cyclically processed and written into the BRAM until receiving read command from computer, and $M$ samples will be transmitted to the computer based on HSC module, as shown in Fig.~\ref{fig2}(c).

In addition, a MSG with two hybrid channels is designed and embedded in the mixed-signal DAQ system, which can generate variable-width pulses based on pulse width modulation (PWM) technique with the time resolution of 8 ns, and sine-type RF signals with the theoretical frequency bandwidth from 0 to 62.5 MHz based on direct digital synthesis (DDS) method\cite{dds1, dds2, dds3}. The digital pulses output from MSG can be utilized for the coarse synchronization between the DAQ system and experimental platform; RF signals with adjustable amplitude and phase can be used for the frequency-modulation (FM) of microwave (MW) frequency, which can also realize the manipulation of nuclear spins in solid-spin systems with the frequency of a few MHz\cite{nuclearspin}.

\section{Calibration of DAQ system}
For the improvement of SAP module in the mixed-signal DAQ system, calibration of amplitude and bias adjustment for analog channels of the system are necessary. The amplitudes of samples from the 14-bit ADC are initially quantified as the original data from 0x0000 to 0x3FFF, which correspond to the decimal values from 0 to 16383. Obviously, this kind of quantization is not suitable for further data processing on computer. To optimize this problem, binary complement format is adopted for encoding the original data from ADC. For example, the highest bit of 14-bit original data acts as sign bit, and binary complement of original data 0x0000 is 01111111111111b, of which the highest bit represents the plus sign and the decimal true form is +8191; similarly the original data 0x3FFF can be encoded as -8192. After the above encoding process, the amplitudes of samples are converted into the decimal values from -8192 to +8191, which correspond to the real voltage range from -1 V to +1 V. In addition, as shown in Fig.~\ref{fig3}(a), it is obvious that there are a few bias noise on both analog channels caused by the defects of hardware. The bias noise adversely affect the accuracy of further data processing. To reduce this impact, correction of bias noise can be implemented based on bias adjustment function designed in hardware. The oscillagram of bias noise with corrections is shown in Fig.~\ref{fig3}(b). Here both oscillagrams in Fig.~\ref{fig3}(a) and Fig.~\ref{fig3}(b) are obtained by the SAP module in the system based on continuous pattern. In fact, environmental temperature mainly affect the bias noise of analog channels in the DAQ system. Generally, the environmental temperature in our experiment is about 24$^{\circ}$C. To determine this impact, after the pre-correction of the bias noise according to the reference under our experimental temperature, we demonstrated the DAQ system in a temperature-controlled container. The environmental temperature in container is heated from 30 to 60$^{\circ}$C by heating resistance and stabilized by a proportional integral derivative (PID) temperature controller with a digital thermometer. The results shown in Table~\ref{tab1} indicate that, with the raising of temperature, the average bias noise approximately linearly increase. As a result, temperature-dependent correction of analog channels is necessary for compensating the change of bias noise.

\begin{table}
	\caption{\label{tab1}Average bias noise of analog channels under different environmental temperatures.}
	\begin{ruledtabular}
		\begin{tabular}{cccc}
			& Temperature ($^{\circ}$C) & Average bias noises (mV) & \\
			\hline
			& 30 & 0.124 & \\
			& 35 & 0.393 & \\
			& 40 & 0.608 & \\
			& 45 & 0.803 & \\
			& 50 & 0.929 & \\
			& 55 & 1.086 & \\
			& 60 & 1.275 & \\
		\end{tabular}
	\end{ruledtabular}
\end{table}

\begin{figure*}
	\includegraphics[width=1.8\columnwidth]{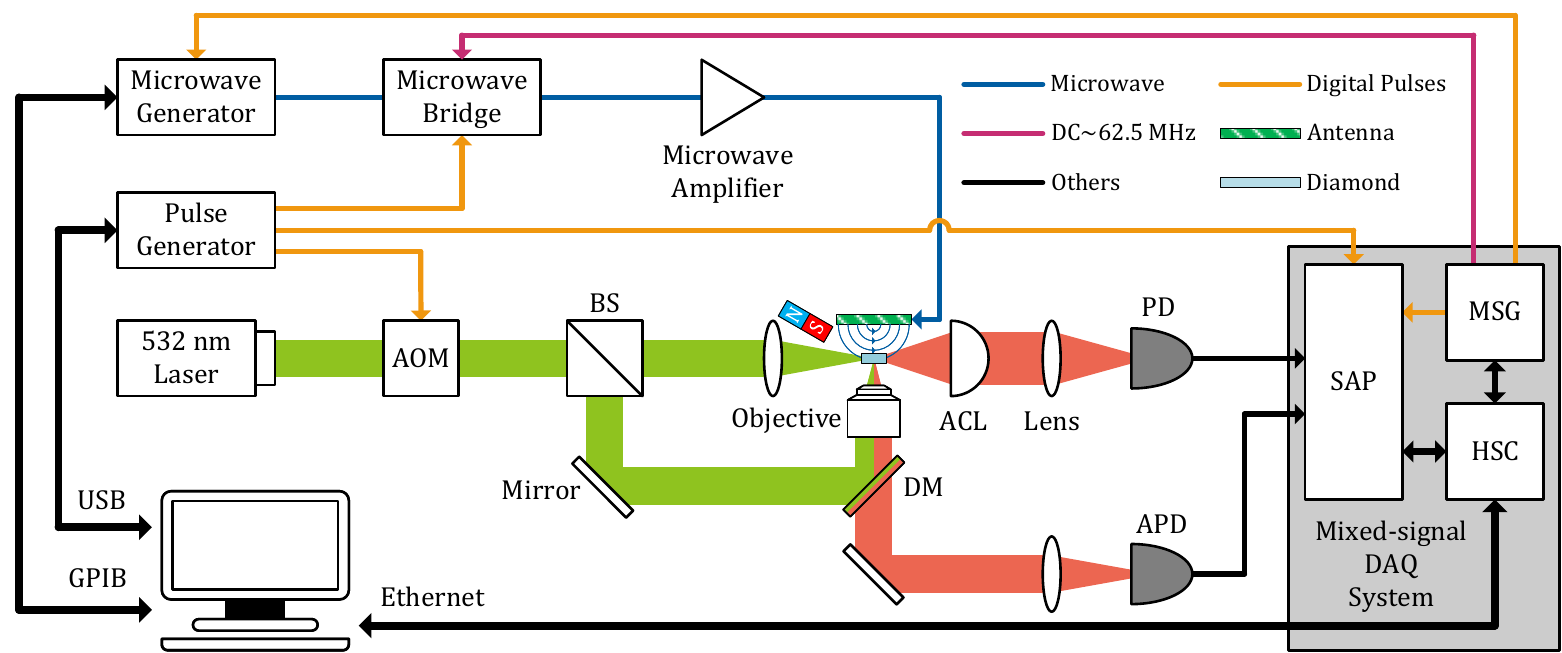}
	\caption{\label{fig4} The overview of purpose-built ODMR spectrometer based on NV centers. ACL, aspheric condenser lens; DM, dichroic mirror. Analog signals from PD and digital signals from APD are sampled and processed by SAP module in synchronization with trigger events from MSG module or external pulse generator. Communications between computer and the DAQ system, MW generator, and pulse generator are established via gigabit ethernet, GPIB, and USB respectively.}
\end{figure*}

As for the MSG module embedded in the reported DAQ system, both analog and digital signals output via hybrid channels are calibrated. As an example, Fig.~\ref{fig3}(c) shows the oscillagram of sine-type RF signals with frequency of 10 MHz and different amplitudes output from MSG. Limited by the drive capability of power supply on system, the maximum amplitude of RF signals is about 2 voltage peak-to-peak (Vpp), which corresponds to the power of about 10 dBm with 50 $\Omega$ load. As the output frequency increases, the output power is mainly subject to slew rate limitation and decreases. For obtaining the actual frequency bandwidth of MSG, power calibration of MSG is indispensable. Fig.~\ref{fig3}(d) indicates that the amplitude of RF signals is reduced by 3 dB at the frequency of 49.57 MHz, and the actual frequency bandwidth of MSG is from 0 to near 50 MHz. Various waveforms like square, triangle, and sawtooth are available as required in MSG module. In addition, the PWM-based digital pulses with adjustable period and duty cycle output from MSG can be used for the coarse synchronization between the DAQ system and experimental platform, the time resolution of which is 8 ns depending on the internal clock rate. Fig.~\ref{fig3}(e) shows the oscillagram of PWM-based digital pulses with different periods and duty cycles. It is obvious that the rising and falling time are quite long and about 10 ns, which is caused by the limited frequency bandwidth of hybrid channels. As a result, these digital pulses are suitable for experiment without the requirement of precise synchronization. In this work, digital pulses with period of 20 ms and duty cycle of 0.8, and sine-type RF signals with the frequency of 0.1 Hertz (Hz) and the amplitude of about 1.72 Vpp, are utilized in quantum experiment of NV centers discussed in next section.

\section{Applications in NV centers}
\subsection{Experimental setup}
We perform the reported mixed-signal DAQ system on a purpose-bulit ODMR spectrometers based on NV centers in diamond, as shown in Fig.~\ref{fig4}. A 532 nm laser modulated by acousto-optic modulator (AOM) is divided into two beams by using a beam splitter (BS). One of them is focused on the diamond to excite NV ensembles by a 5 cm focal-length lens; the other is transmitted to a confocal system and focused  to excite single NV center by using an 100$\times$ oil-immersed objective with the numerical aperture (NA) of 1.35. The fluorescence of NV ensembles is collected by an aspheric condenser lens(Thorlabs ACL25416U-B) with the NA of 0.79, and then detected by PD (Thorlabs PDA36A2); the fluorescence of single NV center is collected by the objective and detected by APD (Excelitas Photon Counting Module). The diamond is mounted on an $\Omega$-type slotline antenna which can generated an alternating magnetic field. The static magnetic field is applied by a columnar neodymium magnet to split the $\left | m_S = \pm 1 \right \rangle$ sublevels of NV centers. The MW signals generated from a vector signal generator (Rohde$\&$Schwarz SMIQ03B) and modulated by MW bridge are amplified by a power amplifier (Minicircuits ZHL-16W-43+) and delivered to the antenna. Precise pulse modulations of lasers and MW signals are realized by using our home-built high-performance pulse generator with the time resolution of 1 ns.

\subsection{General-purpose experiment of NV ensembles}
In order to demonstrate the performance of the reported DAQ system, continuous wave ODMR (cw-ODMR) experiment and Rabi oscillation based on NV ensembles are performed. The energy level of electron spin in NV center is shown in Fig.~\ref{fig4}(a). With the pumping of 532 nm laser, the electron spin is excited from the ground states ($^3$A$_2$) to the excited states ($^3$E). After a few nanoseconds, the electron spin decays back to the ground states and emits the fluorescence from 637 nm to 800 nm. Meantime, with the assistance of intersystem crossing (ISC) transitions, the electron spin on the excited $\left | m_S = \pm 1 \right \rangle$ state can be non-radiatively polarized into the ground $\left | m_S = 0 \right \rangle$ state, which enables the optical readout of the electron spin state. Furthermore, electron spins can be drived from the more fluorescent $\left | m_S = 0 \right \rangle$ ground state into the less fluorescent $\left | m_S = \pm 1 \right \rangle$ state by MW signals, of which the frequency is tuned near the resonance of one of the $\left | m_S = 0 \right \rangle \leftrightarrow \left | m_S = \pm 1 \right \rangle$ transitions.

\begin{figure}
	\includegraphics[width=1\columnwidth]{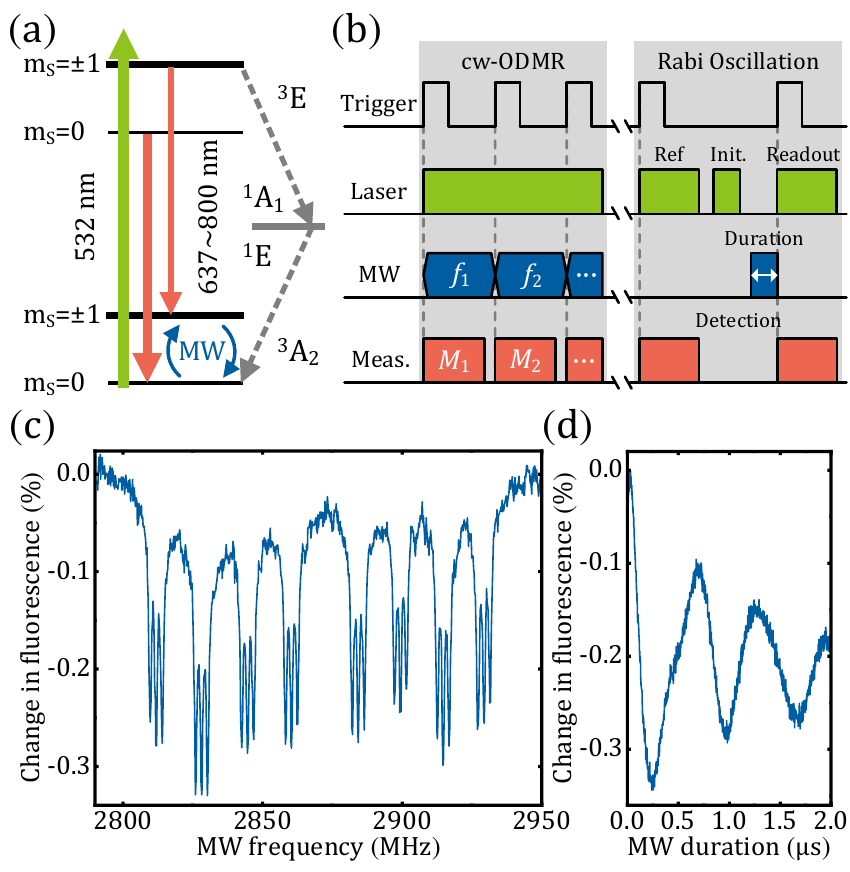}
	\caption{\label{fig5} (a) The schematic energy level of electron spin in NV center. Green and red arrows represent 532 nm laser and fluorescence from 637 nm to 800 nm respectively. Dotted gray arrow path represents the ISC transitions between excited $\left | m_S = \pm 1 \right \rangle$ states and ground $\left | m_S = 0 \right \rangle$ states. $\left | m_S = 0 \right \rangle \leftrightarrow \left | m_S = \pm 1 \right \rangle$ transitions represented by blue arrows are realized by resonant MW signals. (b) Experimental sequence of cw-ODMR and Rabi oscillation experiment based on NV ensembles. The duration of detection windows is fixed at 16 ms. (c) cw-ODMR spectra of NV ensembles. 8 broad resonance peaks are caused by four kinds of NV orientations and 24 tiny resonance peaks are caused by hyperfine interactions of $^{14}$N nuclear spins in NV centers. (d) Rabi oscillation of NV ensembles corresponding to the first broad resonance peak in (c).}
\end{figure}

The experimental sequence of cw-ODMR experiment of NV ensembles is shown in Fig.~\ref{fig5}(b). Continuous laser and MW signals are applied in the whole duration of experiment. Digital pulses with period of 20 ms and duty cycle of 0.8 output from the mixed-signal DAQ system are utilized as trigger events of signal sampling and frequency switching of MW signals. The duration of detection window for each experimental point is 16 ms. Fig.~\ref{fig5}(c) shows the cw-ODMR spectra of NV ensembles. There are 1000 experimental points with the MW frequency range from 2790 MHz to 2950 MHz. 24 accurate resonant frequencies can be extracted from the spectra, which belong to four kinds of NV centers with different orientations: $\left \langle 1 1 1 \right \rangle$, $\left \langle \overline{1} 1 1 \right \rangle$, $\left \langle 1 \overline{1} 1 \right \rangle$, and $\left \langle 1 1 \overline{1} \right \rangle$. Furthermore, pulsed experiment like Rabi oscillation is performed. The experimental sequence of Rabi oscillation is shown in Fig.~\ref{fig5}(b). Here trigger events of signal sampling are provided by pulse generator, and the results of Rabi oscillation are shown in Fig.~\ref{fig5}(d).

\begin{figure}
	\includegraphics[width=1\columnwidth]{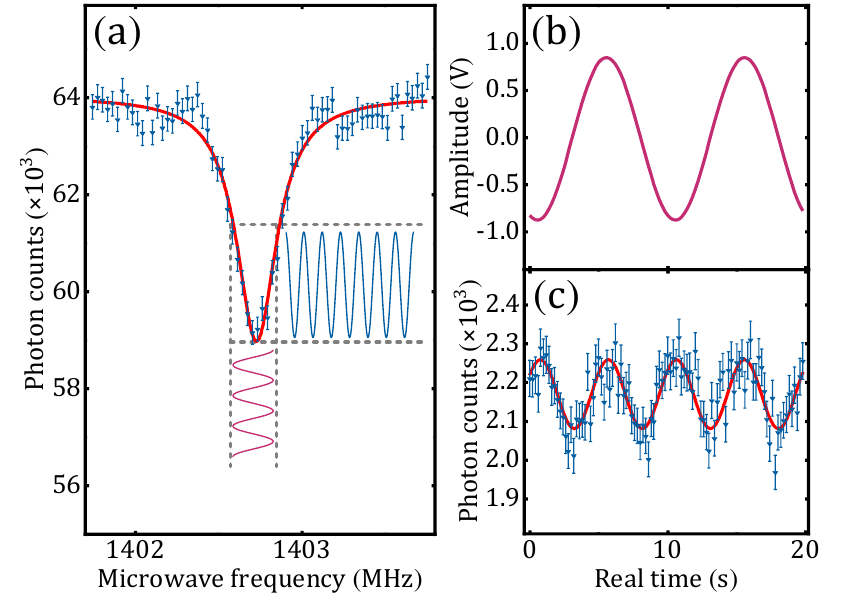}
	\caption{\label{fig6} Lock-in detection experiment of single NV center. (a) Pulsed ODMR spectra of single NV center. The $^{14}$N nuclear spins are optically polarized by applying a static magnetic field about 52 mT\cite{eslac}. (b) The oscillagram of RF signal with the frequency of 0.1 Hz and amplitude of 1.72 Vpp output from the mixed-signal DAQ system. The MW frequency at resonance peak in (a) is FM-modulated by the RF signal. (c) The oscillation of fluorescence caused by FM-modulated MW. Here oscillations in (b) and (c) are in-phase.}
\end{figure}

\subsection{Lock-in detection of single NV center}
The main function of the reported mixed-signal DAQ system is synchronized data acquisition and processing of mixed signals, and it is of great importance for the quantum metrology. For example, the sensitivity of quantum magnetometer based on NV centers can be improved by modulating the MW frequency to combat the 1/f noise from apparatus\cite{NVzongshu}; then the demodulation and analysis of information from quantum megnetometer are depending on the in-phase modulation signals, which is the fundamental of the well-known Lock-in detection. In order to demonstrate this advanced function, the Lock-in detection of single NV center is performed\cite{NVsystem}. Sine-type RF signals generated by MSG with the frequency of 0.1 Hz and the amplitude of 1.72 Vpp are used as the FM modulation signals of MW frequency. The fluorescence of single NV center is detected by APD. Digital signals from APD and the in-phase modulation signals are synchronously acquired by the DAQ system.

Fig.~\ref{fig6}(a) shows the pulsed ODMR spectra of individual electron spin in NV center corresponding to $\left | m_S = 0 \right \rangle \leftrightarrow \left | m_S = -1 \right \rangle$ transition under a static magnetic field about 52 mT. It is obvious that the intensity of fluorescence is minimum when MW is fixed at the resonant frequency. The intensity of fluorescence increases with the increasing MW frequency detuning. Since the MW frequency is modulated by RF signals output from MSG, the intensity of fluorescence oscillates within a certain range determined by amplitude of RF signals. Fig.~\ref{fig6}(b) shows the oscillagram of RF signals obtained by the DAQ system, and Fig.~\ref{fig6}(c) indicates the fluorescence oscillation where the error bars of fluorescence is calculated based on error propagation method\cite{chopped}. According to the fitting results, the frequency of fluorescence oscillation is twice as big as the frequency of RF signals. Note that fluorescence oscillation and RF signals remain in phase, which indicates the outstanding performance on synchronized acquisition and processing of mixed signals.

\section{Conclusion}
We have designed and implemented a mixed-signal DAQ system for ODMR of solid-state spins. Based on the mixed-signal DAQ system, general-purpose experiments of single spin and spin ensembles can be easily performed. Meanwhile, with the advanced functions like synchronized acquisition and processing of mixed signals embedded, the reported DAQ system can be developed as a Lock-in amplifier (LIA), which plays an important role in the field of quantum metrology\cite{mitthesis}. With this DAQ system, advanced experiments like Lock-in detection and synchronized ODMR scanning experiments of NV centers are realized. In addition, extra modules like multiplex signal generator embedded in the DAQ system provide convenience for quantum experiments. Furthermore, benefit from the FPGA-based architecture, the reported DAQ system is reconfigurable, extensible, and low-cost. With further required modules customized and embedded, our DAQ system can be easily extended to other spin-based systems.

\begin{acknowledgments}
This work is supported by the National Key Research and Development Program of China (Grants No. 2018YFA0306600, 2020YFA0309400), the National Natural Science Foundation of China (Grants No. 12174081), and the Fundamental Research Funds for the Central Universities (Grants No. JZ2021HGTB0126, PA2021KCPY0052).
\end{acknowledgments}
~\\
\section*{Author declarations}
\subsection*{Conflict of interest}
The authors have no conflicts to disclose.

\section*{Data availability}
The data that support the findings of this study are available from the corresponding author upon reasonable request.

%\bibliography{reference}% Produces the bibliography via BibTeX.
%

\end{document}